# INTERFACE BETWEEN THE LONG-TERM PROPAGATION AND THE DESTRUCTIVE RE-ENTRY PHASES EXPLOITING THE OVERSHOOT BOUNDARY

**Christian Fusaro[(1)], Mirko Trisolini[(2)], Camilla Colombo[(3)]**

[(1)] MSc Space Engineering graduate, Department of Aerospace Science and Technology, Politecnico di Milano, Via La Masa 34, 20156, Milano, Italy. E-mail: christian.fusaro@mail.polimi.it

[(2)] Postdoc Researcher, Department of Aerospace Science and Technology, Politecnico di Milano, Via La Masa 34, 20156, Milano, Italy. E-mail: mirko.trisolini@polimi.it

[(3)] Associate Professor, Department of Aerospace Science and Technology, Politecnico di Milano, Via La Masa 34, 20156, Milano, Italy. E-mail: camilla.colombo@polimi.it

**ABSTRACT**

In recent years, due to the constant increase of the density of satellites in the space environment, several studies have been focused on the development of active and passive strategies to remove and mitigate space debris.
This work investigates the feasibility of developing a reliable and fast approach to analyze the re-entry of a satellite. The numerical model interfaces the long-term orbit propagation obtained through semi-analytical methods with the atmospheric destructive re-entry phase exploiting the concept of overshoot boundary, highlighting the effect that an early break-off of the solar panels can have on the re-entry prediction. The re-entry of ESA's INTEGRAL mission is chosen as test case to demonstrate the efficiency of the model in producing a complete simulation of the re-entry. The simulation of the destructive re-entry phase is produced using an object-oriented approach, paying attention to the demisability process of the most critical components of the space system.

## 1. INTRODUCTION

Different works proved that the effect of luni-solar perturbations can be exploited to obtain a long-term re-entry of a space system [1][2][3].
This article presents a fast method to analyze the atmospheric re-entry of a satellite enhanced by the effect of orbit perturbations, searching for the correct balance between approximation and precision.
The long-term propagation of the orbital elements is performed through semi-analytical methods implemented in the *PlanODyn* tool available at Politecnico di Milano [4]. The entry conditions, obtained exploiting the overshoot boundary theory [5], are then used in combination with a simplified configuration of the spacecraft to produce a simulation of the destructive re-entry phase using an object-oriented approach. The demisability analysis is performed with the *Phoenix* software [6][7][8].
The model presented can be used for a preliminary identification of the disposal strategies that allow a proper demise of the most critical components of the spacecraft during the re-entry phase.

## 2. LONG-TERM PROPAGATION

The design of the end-of-life strategy is a complex task that requires the propagation of the orbital evolution of the spacecraft over a long time. The orbital dynamics is characterised by the presence of several perturbation effects that increase the computational time required to obtain an exact long-term evolution of the orbital elements. The propagation performed in this work are obtained using a semi-analytical method implemented in the *PlanODyn* tool [4], to reduce the computational cost of the simulation. The model describes the dynamic evolution of the spacecraft exploiting the Lagrange's Planetary Equations and averaged expression of the potential of disturbing effects. All the propagations performed in this work are performed using the above-mentioned tool.

### 2.1 Lagrange's planetary equations

The long-term propagation is computed considering the effects of luni-solar perturbations, of the Earth's oblateness and of the aerodynamic drag. For a perturbed motion, the rates of change of the six orbital elements $(a, e, i, \Omega, \omega, M)$ are expressed by the Lagrange Planetary Equations [9]:

$$\frac{da}{dt} = \frac{2}{na}\frac{\partial R}{\partial M}$$

$$\frac{de}{dt} = \frac{(1-e^2)^{1/2}}{ena^2}\left((1-e^2)^{1/2}\frac{\partial R}{\partial M} - \frac{\partial R}{\partial \omega}\right)$$

$$\frac{di}{dt} = -\frac{1}{na^2 \sin i\,((1-e^2)^{1/2}}\left(\frac{\partial R}{\partial \Omega} - \cos i\frac{\partial R}{\partial \omega}\right) \quad (1)$$

$$\frac{d\Omega}{dt} = \frac{1}{na^2(1-e^2)^{1/2}\sin i}\frac{\partial R}{\partial i}$$

$$\frac{d\omega}{dt} = \frac{(1-e^2)^{1/2}}{ena^2}\frac{\partial R}{\partial e} - \frac{\cos i}{na^2(1-e^2)^{1/2}\sin i}\frac{\partial R}{\partial i}$$

$$\frac{dM}{dt} = n - \frac{2}{na}\frac{\partial R}{\partial a} - \frac{(1-e^2)}{ena^2}\frac{\partial R}{\partial e}$$

where $n$ is the mean motion of the satellite and $R$ is the disturbing function and represents the disturbing part of the potential.

**2.2 Gauss' form of the variational equations**

In case of an impulsive firing the instantaneous variation of the six orbital parameters is computed through Gauss' form of the variational equation. The impulsive firing is characterized by a velocity variation $\delta \boldsymbol{v} = [\delta v_t, \delta v_n, \delta v_h]$ expressed in the velocity reference frame [10].

$$\delta a = \frac{2a^2 v_d}{\mu}\delta v_t$$

$$\delta e = \frac{1}{v_d}\left(2(e + \cos\theta)\delta v_t - \frac{r_d}{a}\sin\theta \delta v_n\right)$$

$$\delta i = \frac{r_d \cos u}{h}\delta v_h$$

$$\delta \Omega = \frac{r_d \sin u}{h \sin i}$$

$$\delta \omega = \frac{1}{ev_d}\left(2\sin\theta \delta v_t + \left(2e + \frac{r_d}{a}\cos\theta\right)\delta v_n\right)$$

$$- \frac{r_d \sin u \cos i}{h \sin i}\delta v_h$$

$$\delta M = -\frac{b}{eav_d}\left(2\left(1 + \frac{e^2 r_d}{p}\right)\sin\theta\, \delta v_t + \frac{r_d}{a}\cos\theta\, \delta v_n\right)$$

(2)

where $r_d$ and $v_d$ are the orbital radius and the velocity at the point where the impulsive firing is performed, $h$ is the angular momentum, $p$ is the semi-latus rectum and $u$ is the argument of latitude defined as $u = \omega + \theta$.
Eqs. 2 are used to compute the instantaneous effect of the disposal maneuver on the orbital elements.

**3. ATMOSPHERIC INTERFACE**

During the design of a strategy that will lead the satellite to an atmospheric re-entry, it is fundamental to produce a prediction of the behavior of the space system during the destructive phase. For this reason, the theoretical concept of overshoot boundary is introduced, with the intent to exploit the method as effective interface between the long-term orbit propagation and the destructive re-entry phases.

**3.1 Entry corridor**

The behavior of the vehicle during the re-entry is highly influenced by the flight conditions at the entry interface. If the entry angle is too shallow, the vehicle may pass through the upper layer of the atmosphere and continue his path in the space environment. On the other side if the entry angle is too steep at the entry interface, the spacecraft will be affected by high mechanical and thermal loads that could exceed the maximum loads allowed by the mission requirements. All the trajectories between these two extreme situations ensure not only that the spacecraft will be captured during its atmospheric passage, but also that the re-entry conditions will not exceed the design limits.
*Figure 1* depicts the two limits that define the range of trajectories that guarantee a safe re-entry of the spacecraft. The overshoot boundary is defined by the maximum periapsis radius which allows an atmospheric capture of the satellite at the first atmospheric passage. Above the overshoot boundary the satellite encounters too small atmospheric drag and will not be captured.
The undershoot boundary is instead related to the maximum deceleration allowed along the re-entry trajectory. If the vehicle enters the atmosphere below this limit, it will experience too much drag: the undershoot boundary is therefore a representation of the border between "safe" and "unsafe" entry. The entry corridor is straightforward defined as the space defined between the two boundaries [5].

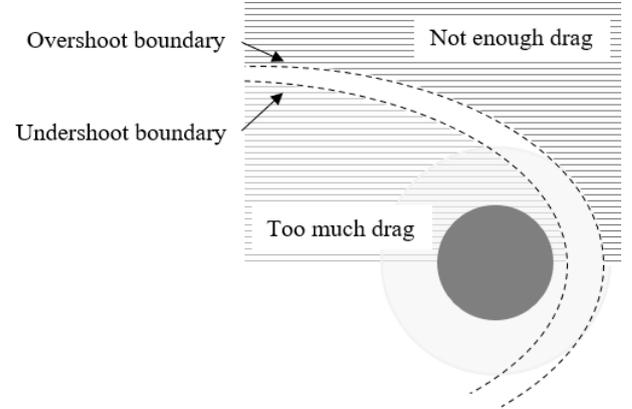

*Figure 1. Entry corridor.*

**3.2 Unified theory**

The unified theory presented in this section follows the approach adopted in Hicks [5], obtained following the method described by Vinh [11]. The main issues that affect the analysis of a re-entry strategy are related to the impossibility of defining an exact border between the deep space and the Earth's atmosphere and to the difficulty of identifying the flight conditions that will allow an atmospheric capture of the spacecraft. The overshoot theory allows us to drop the assumption of a fixed altitude for the atmospheric interface, solving the problem of the prediction of the re-entry conditions from a point of view of the deceleration experienced by the satellite along its trajectory. The approach presented in this section allows us to include the drag effect in the long-term propagation, producing a more refined estimation of the possible entry conditions.

The overshoot boundary is computed selecting an arbitrary value $f$ for the ratio between the deceleration and the gravity acceleration. When the ratio reaches the value $f$ then the atmospheric entry occurs. Along the overshoot boundary the ratio defined above should therefore never drop below $f$. The deceleration ratio is somewhat arbitrary. According to Vinh's indication it is set here equal to 0.05. The overshoot boundary theory is based on the definition of two adimensional variables called "modified" Chapman variables [5]

$$F = \frac{\rho(r) S C_D}{2m}\sqrt{\frac{r}{\beta}} = Z \tag{3}$$

$$u = \frac{v^2 \cos^2 \gamma}{g(r) r} \tag{4}$$

where $\rho$ is the atmospheric density at distance $r$. $S$ is the cross-sectional area of the spacecraft, $C_D$ is the drag coefficient, $m$ is the mass of the vehicle, $\beta$ is the inverse of the atmospheric scale height, $g$ is the gravitational acceleration and $v$ and $\gamma$ are the velocity and the flight path angle of the vehicle, respectively.

The procedure to build the overshoot boundary is obtained as follows:

- *STEP 1*: Choose a value for $Z_*$. The value defines the point of critical deceleration, where $(a/g_0) = f$.

- *STEP 2*: Solve equations

$$2\sqrt{\overline{\beta r}} Z_* + (\overline{\beta r} - 1) \sin \gamma_* + \frac{2 \sin \gamma_* \cos^2 \gamma_*}{u_*} = 0 \tag{5}$$

$$\left(\frac{a_{decel}}{g_0}\right)_* = \frac{Z_* u_* \sqrt{\overline{\beta r}}}{\cos^2 \gamma_*} + \sqrt{1 + \left(\frac{C_L}{C_D}\right)^2} = f$$

for the adimensional velocity $u_*$ and the flight-path angle $\gamma_*$, where $C_L$ is the lift coefficient of the spacecraft and $\overline{\beta r}$ is an average value characterizing the atmosphere considered. The solution defined by $Z_*$, $u_*$ and $\gamma_*$ represents a point of minimum deceleration on the trajectory.

- *STEP 3*: Recover the entry conditions

$$\frac{dZ}{ds} = -\overline{\beta r} Z \tan \gamma \tag{6}$$

$$\frac{du}{ds} = \frac{2Zu\sqrt{\overline{\beta r}}}{\cos \gamma}\left(1 + \frac{C_L}{C_D}\cos \sigma \tan \gamma + \frac{\sin \gamma}{2Z\overline{\beta r}}\right) \tag{7}$$

$$\frac{d\gamma}{ds} = \frac{Z\sqrt{\overline{\beta r}}}{\cos^2 \gamma}\left[\frac{C_L}{C_D}\cos \sigma + \frac{\cos \gamma}{Z\overline{\beta r}}\left(1 - \frac{\cos^2 \gamma}{u}\right)\right] \tag{8}$$

The equations are expressed using the arc-length $s$ traveled by the spacecraft. The entry conditions are calculated integrating Eqs. 6-8 "backwards" along $s$ until the condition $(a_{decel}/g_0) = f$ is obtained again. The variable $\sigma$ represents bank angle, which is kept constant during the motion of the spacecraft. The equations reported above represent a two-point boundary value problem, where boundary conditions are defined as

$$Z = Z_*, u = u_*, \gamma = \gamma_* \text{ at } s = s_*$$

and

$$\frac{Z_e u_e \sqrt{\overline{\beta r}}}{\cos^2 \gamma_e} + \sqrt{1 + \left(\frac{C_L}{C_D}\right)^2} = f \text{ at } s = 0$$

- *STEP 4*: Solve equations

$$\frac{u_e^2}{\cos^2 \gamma_e} - 2u_e = u_p^2 - 2u_p$$

$$Z_p = Z_e \sqrt{\frac{u_e}{u_p}} exp\left[\overline{\beta r}\left(1 - \frac{u_e}{u_p}\right)\right] \tag{9}$$

simultaneously to find $F_{p_{over}} = Z_{p_{over}}$

Repeating the procedure presented above for different values of $Z_*$ allows us to obtain the expression of the periapsis parameter $Z_p$ as a function of the adimensional velocity $u_e$.

*Figure 2* shows an overshoot boundary computed for ballistic entry $(C_L/C_D = 0)$ into Earth's atmosphere $(\overline{\beta r} = 900)$, considering a deceleration ratio $f = 0.05$. The expression of the overshoot boundary, which is obtained without using any information about the satellite configuration, is based on adimensional variables; the boundary can therefore be pre-computed and applied for the re-entry prediction of missions characterized by the same average atmosphere property $(\overline{\beta r}$, i.e. same planet) and the same lift-to-drag ratio $(C_L/C_D)$.

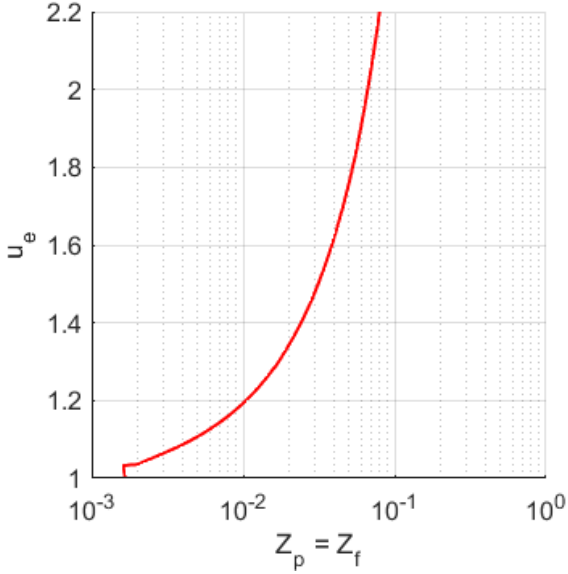
*Figure 2. Overshoot boundary.*

## 4. The re-entry of INTEGRAL

The unified theory is here used to predict the re-entry conditions that characterize the re-entry from highly elliptical orbits. An optimized disposal strategy of INTEGRAL is used as test case [1].

### 4.1 Characteristics of the spacecraft

The re-entry prediction is performed for an area-to-mass ratio associated to an average value for the cross-section according to the assumption of random tumbling of the spacecraft. The average cross-section is computed as

$$S = \frac{1}{3}(A_{SP} + WH + LH + LW) \qquad (10)$$

where $A_{SP}$ is the solar panels area and W, H and L are are the dimensions of the spacecraft in the orbit configuration.
*Table 1* summarizes the characteristics of INTEGRAL.

| | |
|---|---|
| Solar panels area | 21.9 m² |
| L | 2.175 m |
| W | 3.138 m |
| H | 4.939 m |
| Dry mass | 3414 kg |
| Drag coefficient | 2.2 |

*Table 1. Characteristics of INTEGRAL's configuration.*

### 4.2 Disposal strategy

The disposal strategy considered was obtained by Colombo et al. [1] through an optimization procedure. The strategy design demonstrated that the re-entry of the spacecraft can be achieved with a low velocity variation and exploiting luni-solar perturbations. The re-entry of the satellite is obtained with a single-maneuver that leads the satellite, affected by luni-solar perturbations, to a long-term uncontrolled re-entry.

The maneuver considered is characterized by impulsive velocity change performed on 08/08/2014 at 9:00 A.M. The velocity variation is completely defined through four parameters: the in-plane and the out-of plane angles ($\alpha$ and $\beta$) of the velocity variation expressed in the velocity reference frame, the magnitude of the velocity change $\Delta v$ and the true anomaly $\theta$ at the point where the maneuver is performed.
*Table 2* summarizes the parameters that describe the disposal maneuver.

| Parameter | Value |
|---|---|
| $\Delta v$ | 26.26 m/s |
| $\alpha$ | 173.4 deg |
| $\beta$ | 35.4 deg |
| $\theta$ | 45.1 deg |

*Table 2. Maneuver parameters*

### 4.3 Methodology

The unified theory presented in Section 3.2 is used to estimate the conditions that correspond to an atmospheric re-entry and to identify therefore the time instant when the entry conditions are met.
The values of $Z_p$, $Z_e$ and $u_e$ obtained during the computation of the overshoot boundary are used to obtain an explicit expression of the variables at the entry interface as function of $Z_p$.

$$Z_e = Z_e(Z_p) \qquad (11)$$

$$u_e = u_e(Z_p) \qquad (12)$$

The expressions represented by Eqs. 11-12 are obtained through a polynomial interpolation of the set of data obtained during the computation of the overshoot boundary.
During the long-term propagation, at each time step, the following procedure is performed to check if the re-entry conditions are met.

- *STEP 1*: The value of $Z_{p_d}$ of the disposal orbit must be computed solving Eq. 3.

- *STEP 2*: From $Z_{p_d}$ the respective values of $Z_e$ and $u_e$ are recovered through Eqs. 11-12.

- *STEP 3*: Compute the radius at the entry interface $r_e$ solving Eq. 3 for the value of $Z_e$ recovered in the previous step.

- *STEP 4*: Find the true anomaly at the entry interface as

$$\theta_{entry} = arccos\left(\left(\frac{1}{e}\right)\left(\left(\frac{e}{r_e}\right)(1-e^2)-1\right)\right) \quad (13)$$

- STEP 5: Find the velocity $v_{e_d}$ and the flight-path angle $\gamma_{e_d}$ of the disposal at the overshoot boundary

$$v_{e_d} = \sqrt{\mu\left(\frac{2}{r_e}-\frac{1}{a_d}\right)} \quad (14)$$

$$\gamma_{e_d} = \arctan\left(\frac{e\sin\vartheta_{entry}}{1+e\cos\vartheta_{entry}}\right) \quad (15)$$

- STEP 6: Compute the adimensional entry velocity $u_{e_d}$ using $v_{e_d}$ and $\gamma_{e_d}$

$$u_{e_d} = \frac{v_{e_d}^2 \cos^2\gamma_{e_d}}{gr} \quad (16)$$

- STEP 7: If $u_{e_d} < u_{e_b}$ the orbit meets the conditions that can lead to an atmospheric entry.

## 5. RE-ENTRY ESTIMATION

The methodology presented in section 4.3 is used in the analysis of the disposal strategy of INTEGRAL to estimate the range of the possible entry conditions.

### 5.1 Uncertainty analysis

A set of 500 possible re-entry trajectories is obtained introducing uncertainties for the magnitude of the velocity variation and for the in- and out-of-plane angles of the maneuver. The uncertainties on the in-plane and out-of plane angles are selected according to the pointing and attitude requirements reported in INTEGRAL's manual [12]. The results presented in the next section are obtained considering a normal distribution around the nominal values considering a standard deviation of 2% for the velocity magnitude and of 0.25 degrees for both maneuver angles.
*Figure 3* shows the distribution of the set of $\Delta v$, $\alpha$ and $\beta$ values considered in the analysis.
The propagation of the 500 test cases is firstly performed without the contribution of the atmospheric drag, until a target perigee altitude of 50 km is reached. The re-entry conditions are then obtained retrieving the flight conditions at an altitude of 120 km [1]. The same 500 initial conditions are used to perform a long-term propagation of the orbit including the drag effect until the entry conditions, defined by the overshoot boundary, are met.

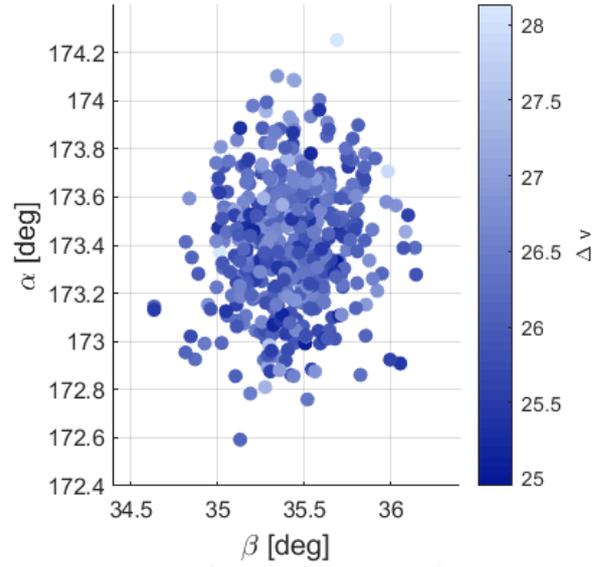

*Figure 3. Distribution of maneuver characteristics affected by uncertainties: 500 test cases.*

### 5.2 Solar-panels break-off

The behavior of the spacecraft during the re-entry is influenced by the area-to-mass ratio of the vehicle.
The re-entry trajectory of the complete vehicle, considering the presence of solar panels, is therefore different from the trajectory of the main body.
*Figure 4* shows the effect of the cross-section on the expression of the overshoot boundary. The trend of the boundary is represented for two values of the area-to-mass ratio, related to the configuration of INTEGRAL with (blue line) and without solar panels (orange line). The two values are computed according to Eq. 10.

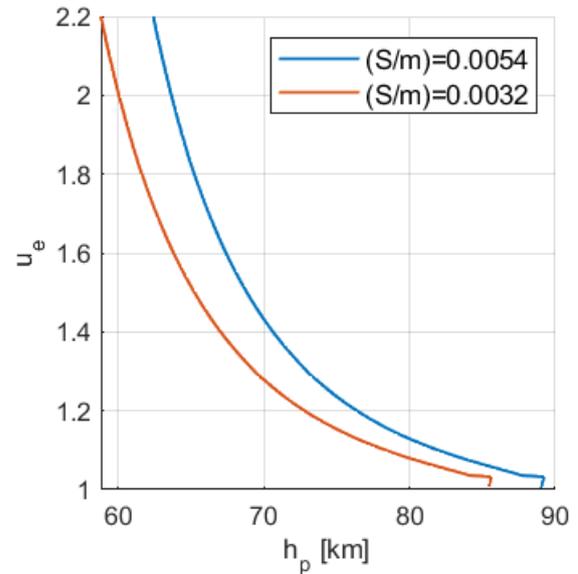

*Figure 4. Overshoot boundary and area-to-mass ratio influence.*

As shown by the overshoot boundaries depicted in *Figure 4*, a higher mean cross-section area is related to higher perigee altitudes of the entry interface; a higher area-to-mass ratio predicts therefore an earlier re-entry of the spacecraft. A preliminary estimation of the re-entry of the spacecraft showed that the re-entry conditions are usually met in the range of altitudes where generally a break-off of the solar panels is verified; therefore, a study of the influence of a variation of the cross-section on the re-entry prediction, is performed. Once the solar-panels break-off altitude is reached, the area-to-mass ratio used in the long-term propagation and in the overshoot boundary theory is reduced to an averaged value characterizing only the main body of the spacecraft. A fixed altitude of 95 km is selected to model the break-off of the solar panels, according to the typical break-up altitudes used in SESAM and other object-oriented models [13].

The set of 500 possible re-entry trajectories obtained introducing uncertainties in the disposal maneuver was analyzed with the overshoot theory to obtain an estimation of the possible re-entry conditions and to examine the effect of a variation of the cross-section on the re-entry prediction. The results of the computation showed that only the 10% of the re-entry trajectories are characterized by a perigee passage below 95 km of altitude that leads to a break-off of the solar panels before the entry conditions are met. In the other cases instead, the re-entry conditions are achieved before a detachment of the solar panels occurs.

*Table 3* and *Table 4* summarize the mean and the standard deviation of the entry conditions obtained in the analysis considering a fixed interface approach and exploiting the overshoot theory, respectively. Considering a fixed interface altitude, and neglecting therefore the drag effect, the long-term propagation produces a set of possible entry conditions characterized by a standard deviation close to zero. Exploiting the overshoot theory and including therefore the drag effect in the long-term propagation, the possible entry conditions spread over a wider range. *Table 4* highlights that, if an early break-off of the solar-panels is present, the entry flight-path angle and the entry velocity assume lower values with respect to the disposal trajectories that don't present the detachment of the solar panels before the re-entry.

| Fixed interface (120 km) | |
|---|---|
| Number of cases | 500/500 |
| mean $v_e$ | 10.86 km/s |
| $v_e$ std. deviation | ~0.00 |
| mean $\gamma_e$ | -5.84 deg |
| $\gamma_e$ std. deviation | ~0.00 |
| mean $h_e$ | 120 km |
| $h_e$ std. deviation | ~0.00 |

*Table 3. Average entry conditions obtained considering a fixed interface altitude.*

| | Overshoot theory | |
|---|---|---|
| | No solar panels break-off | Solar panels break-off |
| Number of cases | 450/500 | 50/500 |
| mean $v_e$ | 10.36 km/s | 9.52 km/s |
| $v_e$ std. deviation | 0.26 | 0.66 |
| mean $\gamma_e$ | -3.68 deg | -2.74 deg |
| $\gamma_e$ std. deviation | 0.14 | 0.72 |
| mean $h_e$ | 99.66 km | 94.12 km |
| $h_e$ std. deviation | 0.54 | 0.91 |

*Table 4. Average entry conditions obtained exploiting the overshoot theory.*

*Figure 5* shows the average number of low altitude passages (below 120 km) performed by the spacecraft before the entry conditions are reached, as function of the perigee altitude of the re-entry orbit. The trajectories that are associated to an early break-off of the solar panels are also characterized by a higher number of atmospheric passages before the re-entry. A higher number of low altitude passages leads to entry orbits with lower eccentricities and higher pericenter altitudes.

It is important to note that the drag effect acting on the spacecraft, responsible of the circularization of the orbit, could also produce thermal and mechanical loads high enough to produce an early fragmentation of the vehicle, implying a possible release of components of the vehicle in orbit [14][15]. Re-entries affected by a higher circularization should therefore be avoided.

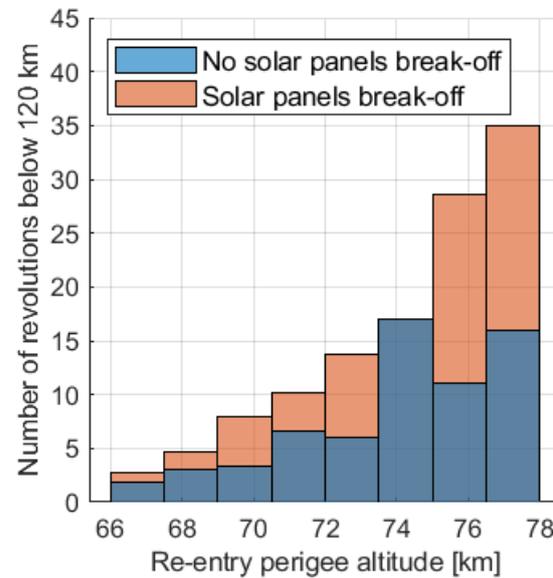

*Figure 5. Average number of revolutions with low altitudes passages (below 120 km).*

## 6. ATMOSPHERIC FLIGHT

### 6.1 Entry trajectory

The atmospheric flight phase of the re-entry trajectory of INTEGRAL is described assuming planar motion and no

lift in a non-rotating environment according to the following equations [16]

$$\frac{dv}{dt} = -\frac{\rho(h)}{2B}v^2 + g(h)\sin\varphi$$

$$v\frac{d\varphi}{dt} = \cos\varphi\left(g(h) - \frac{v^2}{R_E + h}\right) \quad (17)$$

$$\frac{dh}{dt} = -v\sin\varphi$$

$$\frac{dx}{dt} = v\cos\varphi$$

where $B$ is the ballistic coefficient, $v$ is the inertial velocity magnitude, $\varphi$ is the opposite of the flight-path angle, $h$ is the altitude of the spacecraft and $x$ is the downrange. The ballistic coefficient is defined as

$$B = \frac{m}{C_D S} \quad (18)$$

*Figure 6* shows the effect of a maneuver uncertainty on the entry trajectory. The results are compared with entry trajectory computed in a previous work (black line), where the entry conditions were obtained propagating the orbital evolution until a target perigee altitude of 50 km is reached [1]. The entry conditions are then obtained retrieving the flight conditions at an altitude of 120 km. The overshoot theory predicts instead a slower re-entry with lower entry flight path angles.

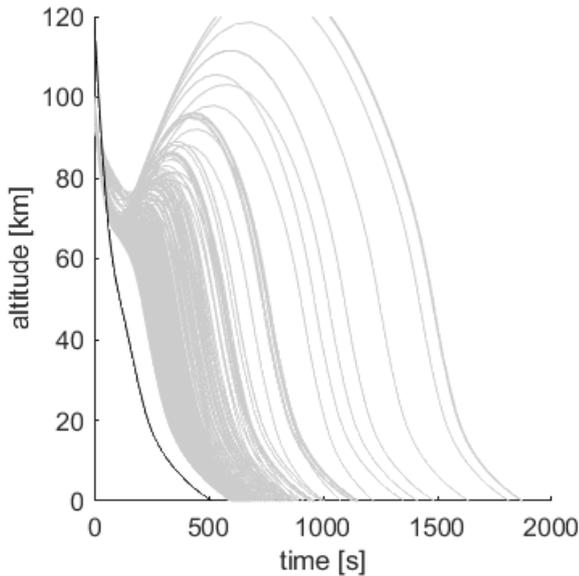

*Figure 6. Predicted re-entry trajectories.*

The re-entry from highly elliptical orbit is characterized by high energy and is predicted to occur close to the perigee, at low entry flight-path angles. The trajectories can therefore present an initial bounce of the spacecraft on the upper layer of the atmosphere.

## 6.2 Mechanical and thermal loads

A preliminary estimation of the mechanical and thermal loads acting on the spacecraft during the atmospheric re-entry is obtained solving the following equations [17]

$$n = \frac{1}{g_0}\left|-\frac{D}{m} + g(h)\sin\varphi\right| \quad (19)$$

$$q = C_D \frac{\rho(h)v^3}{2} \quad (20)$$

where $g_0$ is the ground level gravitational acceleration and $D$ is defined as

$$D = \frac{1}{2}\rho(h)v^2 C_D S \quad (21)$$

*Figure 7* and *Figure 8* show respectively the trend of the mechanical load and of the thermal flux density affecting the spacecraft during the re-entry phase. The main difference between the two approaches used to identify the entry conditions is highlighted by the trend of the loads during the re-entry. Simulations performed exploiting the overshoot boundary theory produce loads that reach peak values at two different altitudes.

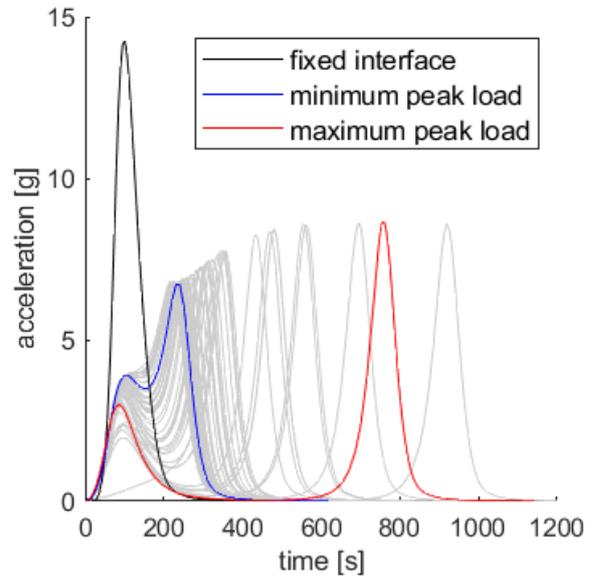

*Figure 7. Predicted evolution of the mechanical load acting on INTEGRAL.*

The entry conditions computed with the overshoot theory predict a slower re-entry that implies a lower altitude for the maximum acceleration. The trend shows that the uncertainties considered for the disposal maneuver produce an uncertainty of around 2 g on the peak acceleration experienced by the vehicle during the re-entry.

The thermal flux density, depicted in *Figure 8*, is characterized by a first maximum at an altitude around 70 km of and a second global maximum at lower altitudes. The method used to compute the entry conditions is therefore highly influencing the estimation of the trend of the mechanical and thermal loads during the re-entry phase.

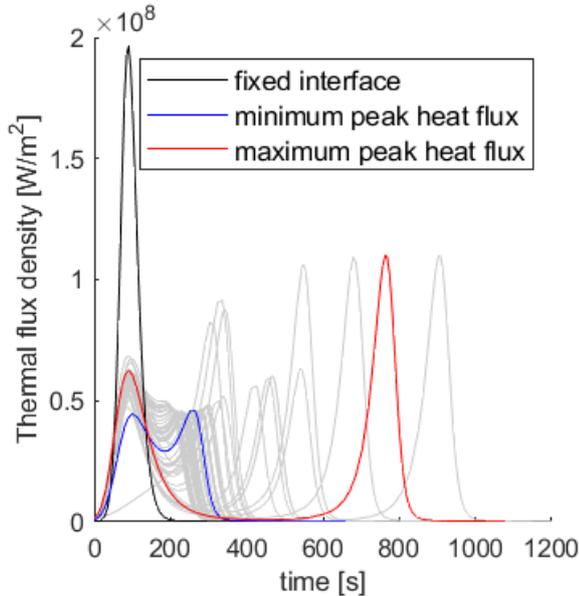

*Figure 8. Predicted evolution of the thermal flux density acting on INTEGRAL.*

## 7. DEMISABILITY ANALYSIS

The entry conditions obtained exploiting the overshoot theory are used as input in the *Phoenix* software in combination with a simplified configuration of the spacecraft, produced according to the user's manual of INTEGRAL [12]. The tool implements an object-oriented method to perform the demisability analysis of the spacecraft. The spacecraft's configuration used in the *Phoenix* tool is reported in *Table A.1*. The model used in the analysis accounts for about 600 kg of the spacecraft mass.

The demisability analysis was performed focusing on the possible entry trajectories that are not affected by an early break-off of the solar-panels, as reported in *Table 4*.

*Figure 9* shows the correlation between the entry conditions and the Liquid Mass Fraction (LMF) predicted for the re-entry. Generally, re-entries with low entry flight-path angles and low entry velocities are affected by a lower mechanical and thermal load during the atmospheric flight phase and are therefore characterized by a higher landing mass.

Note that, despite the demisability analysis is performed on a simplified configuration of the spacecraft, the model can provide a fast estimation of possible entry trajectories that are characterized by a better demise of the most critical components.

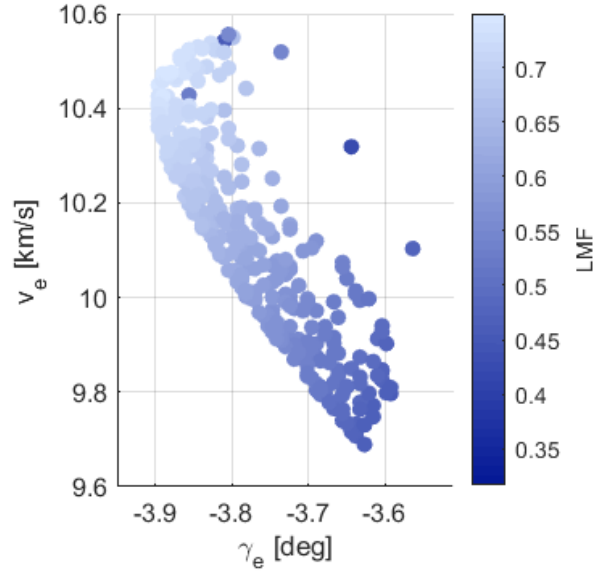

*Figure 9. LMF predicted by Phoenix.*

## 8. CONCLUSIONS

The approach presented in this work defines a fast method for the analysis of the atmospheric re-entry of a satellite enhanced by the effect of orbit perturbations. The model obtained interfacing the semi-analytical orbit propagator and the object-oriented software allows us to obtain an estimation of a disposal strategy in few seconds. The low computational cost makes the model suitable for use in optimization procedures. The method can therefore be used to perform the optimization of the disposal strategy based on the prediction of the risks associated to the re-entry, the residual mass fraction of the satellite impacting on ground and other parameters that characterize the re-entry phase. The model obtained coupling the long-term propagation and an object-oriented model can so be used to identify the disposal maneuvers that allow a proper demise of the most critical components of the spacecraft during the re-entry phase.

## 9. AKNOWLEDGEMENTS

This work was funded by the European Research Council (ERC) under the European Union's Horizon 2020 research and innovation programme (grant agreement No 679086 - COMPASS).

## 11. APPENDIX A

This section reports the simplified configuration of INTEGRAL used as input in the *Phoenix* software. The configuration is summarized in *Table A.1*. For each component the table reports the geometrical shape, the material, the mass (thermal mass m and aerodynamic mass $m_{aero}$), the dimensions (l, r/w and h), the thickness ($t_s$) and the number of components considered ($n_c$).

The parent parameter is used to specify the relation between two components. The parent ID identifies the component that contains the one specified.

The model used in this work accounts for about 600 kg of the spacecraft mass, out of the 3414 kg of its total dry mass. Due to the difficulties of producing a detailed model of the spacecraft, the focus is on the most critical and best-known components.

| PART | ID | PARENT | SHAPE | MATERIAL | m [kg] | l [m] | r/w [m] | h [m] | $t_s$ [mm] | $m_{aero}$ [kg] | $n_c$ |
|---|---|---|---|---|---|---|---|---|---|---|---|
| INTEGRAL PARENT | 0 | \ | Box | Al 6061-T6 | - | 2.175 | 3.138 | 4.939 | 3 | - | 1 |
| Solar panels | 1 | 0 | Flat plate | Al 6061-T6 | 40 | 11.34 | 1.94 | - | - | - | 1 |
| **SERVICE MODULE** | | | | | | | | | | | |
| RW Box 1 | 2 | 0 | Box | Al 6061-T6 | - | 0.35 | 0.35 | 0.35 | 2 | - | 1 |
| RW Box 2 | 3 | 0 | Box | Al 6061-T6 | - | 0.35 | 0.35 | 0.35 | 2 | - | 1 |
| Rw 1 | 4 | 2 | Cylinder | SS AISI-316 | 2.9 | 0.09 | 0.1 | - | - | - | 4 |
| Rw 2 | 5 | 3 | Cylinder | SS AISI-316 | 2.9 | 0.09 | 0.1 | - | - | - | 4 |
| Battery Box 1 | 6 | 0 | Box | Al 6061-T6 | - | 0.45 | 0.45 | 0.25 | 2 | 5 | 1 |
| Battery Box 2 | 7 | 0 | Box | Al 6061-T6 | - | 0.45 | 0.45 | 0.25 | 2 | 5 | 1 |
| Battery cell 1 | 8 | 6 | Box | Al 6061-T6 | 0.135 | 0.07 | 0.06 | 0.018 | - | - | 80 |
| Battery cell 2 | 9 | 7 | box | Al 6061-T6 | 0.135 | 0.07 | 0.06 | 0.018 | - | - | 80 |
| Tank | 10 | 0 | Sphere | Ti 6A14V | - | - | 0.375 | - | 2 | - | 4 |
| Star tracker | 11 | 0 | Cylinder | Al 7075-T6 | 0.47 | 0.188 | 0.06 | - | - | - | 2 |
| OBDH | 12 | 0 | Box | Al 6061-T6 | - | 0.3 | 0.3 | 0.3 | 2 | 15 | 2 |
| Solar array mech | 13 | 0 | Box | SS AISI-316 | 1.5 | 0.12 | 0.11 | 0.07 | - | - | 2 |
| Thruster | 14 | 0 | Cylinder | Inconel-600 | - | 0.12 | 0.05 | - | - | - | 12 |
| Thruster EBox | 15 | 0 | Box | Al 6061-T6 | - | 0.12 | 0.12 | 0.12 | 2 | 5 | 2 |
| Antenna | 16 | 0 | Cylinder | Al 7075-T6 | 35 | 0.01 | 0.75 | - | - | - | 2 |
| Antenna Mech | 17 | 0 | Box | SS AISI-316 | 1.8 | 0.15 | 0.11 | 0.07 | - | - | 2 |
| **PAYLOAD MODULE** | | | | | | | | | | | |
| SPI | | | | | | | | | | | |
| Camera box | 18 | 0 | Box | Beryllium | - | 0.4 | 0.4 | 0.1 | 2 | - | 1 |
| SPI mask | 19 | 0 | Flat plate | Tungsten | - | 0.36 | 0.36 | - | 2 | - | 1 |
| Support plate | 20 | 0 | Flat plate | Ti 6Al4V | 1.8 | 0.45 | 0.45 | - | - | - | 1 |
| IBIS | | | | | | | | | | | |
| Coded Mask | 21 | 0 | Flat plate | Tungsten | 154 | 0.5 | 1 | - | - | - | 1 |
| Ibis box | 22 | 0 | Box | Beryllium | - | 0.5 | 1 | 1 | 2 | - | 1 |

*Table A.1. Simplified configuration of INTEGRAL used in the demisability analysis performed with the Phoenix software.*